# Asymmetry of tensile *vs.* compressive elasticity and permeability contributes to the regulation of exchanges in collagen gels

Jean Cacheux,[1,2] Jose Ordonez-Miranda,[1,2] Aurélien Bancaud,[1,2,3]* Laurent Jalabert,[1,2] Masahiro Nomura,[1,2] Yukiko T. Matsunaga[1,2]†

[1] Institute of Industrial Science, The University of Tokyo, Tokyo 153-8505, Japan.
[2] LIMMS, CNRS-IIS IRL 2820, The University of Tokyo, Tokyo 153-8505, Japan.
[3] LAAS-CNRS, Université de Toulouse, CNRS, Toulouse, France.
*abancaud@laas.fr; †mat@iis.u-tokyo.ac.jp

**Abstract**
The Starling principle describes exchanges in tissues based on the balance of hydrostatic and osmotic flows. This balance neglects the coupling between mechanics and hydrodynamics, a questionable assumption in strained elastic tissues due to intravascular pressure. Here, we measure the elasticity and permeability of collagen gels under tensile and compressive stress via the comparison of the temporal evolution of pressure in an air cavity sealed at the outlet of a collagen slab with an analytical kinetic model. We observe a drop in the permeability and enhanced strain-stiffening of native collagen gels under compression, both effects being essentially lost after chemical cross-linking. Further, we prove that this asymmetric response accounts for the accumulation of compressive stress upon sinusoidal fluid injection, which modulates the material's permeability. Our results thus show that the properties of collagen gels regulate molecular exchanges and could help understand drug transport in tissues.

**Teaser**
Collagen gels contribute to tissue homeostasis through the regulation of the permeability under sinusoidal fluid injection.

**Introduction**
Tissues constitute porous solid matrices traversed by fluid flows that convey the oxygen and nutrients from blood to cells and remove cellular waste products. The Starling principle describes fluid movements between blood and tissues as a balance between the hydrostatic and oncotic pressures (protein component of the osmotic pressure) (*1*). This model has recently been revised to include the effects of lymphatic systems enabling the fluid drainage in tissues (*2*). While hydrodynamic laws suffice to describe this process in rigid porous materials, the change of filtration flows associated to mechanical deformation of soft liquid-filled matrices remains essentially overlooked in physiological transport models. The interaction between fluid flow and solid deformation is described by the poroelastic mechanics, which couples Darcy's law with Terzaghi's effective stress and linear elasticity (*3*). The filtration of liquid in a soft and porous tube differs from that of a hard material, as predicted by analytical models which show the confinement of the flow within a thin layer near the tissue surface (*4*, *5*). These models rely on mechanical (elastic modulus) and hydraulic (permeability) parameters of poroelastic materials, and were recently refined to integrate the effects of nonlinear mechanics and the deformation-dependent permeability (*6*, *7*). To the best of our knowledge, however, linear and non-linear models have not yet been used to characterize the poromechanical response of the main constituents of tissues.

Collagen gels are extensively used as cell culture scaffolds due to their low immunogenicity and versatile elasticity, associated to a shear modulus spanning 1 to 1000 Pa by tuning the gel concentration and reticulation temperature (*8–10*). They recapitulate the constant reorganization of tissues *in vivo* as a consequence of cellular traction forces (*11*, *12*). The permeability of these materials, which ensures transport of solutes to promote cell growth, is also one of their assets (*13*). However, the permeability of collagen gels remains poorly estimated with data spanning two orders of magnitude from $10^{-12}$ to $10^{-14}$ m$^2$ in unstrained conditions (*14–16*). This discrepancy may be resolved by considering the compressive stress that lowers the permeability, but the high load of 700 Pa, which is required to match the different studies, raises questions on whether the coupling between mechanics and hydrodynamics has been accurately described in this material.

In this work, we present a contact-free technology to measure the permeability and elastic modulus of reconstituted collagen networks under compressive and tensile stresses by means of a pressure sensor connected to an air cavity placed at the outlet of the material (Fig. 1A). At the inlet, we apply tunable hydrostatic pressure loads, which induce solid and fluid displacements that compress the air in the cavity (Fig. 1B). The permeability and elasticity of the material are inferred from fitting the measured intracavity pressure to an analytical model that we derive and validate with finite element simulations. The obtained results uncover the asymmetry of the permeation and elastic properties of collagen gels for tensile and compressive stresses. Further, by monitoring the response of this material to pulsatile pressure actuation, we find out that a periodic signal with zero mean value can generate a build-up in bulk compressive stress that in turn tunes the gel permeability.

**Results**

*I. Poroelastic model of the pressure in the air cavity*
Let us consider a homogeneous poroelastic material, in which the fluid and solid displacements are unidirectional between its inlet ($x = 0$) an outlet ($x = H$), as shown in Fig. 1A. According to Darcy's law, the permeation flow associated to a velocity $W$ is proportional to the pressure field $P(x, t)$ gradient, as follows

$$\frac{\partial P}{\partial x} = -\frac{\mu}{\kappa}W, \quad (1)$$

where $\mu$ is the fluid viscosity and $\kappa$ the material permeability. The pressure gradient in Eq. (1) is related to the deformation field $U(x,t)$ of the solid by the balance of mechanical equilibrium (17)

$$M\frac{\partial^2 U}{\partial x^2} = \frac{\partial P}{\partial x}, \quad (2)$$

with $M$ being the elastic modulus of the considered isotropic homogeneous material. Considering that the total flow rate of the liquid $W$ and solid $\partial U/\partial t$ are conservative ($W + \partial U/\partial t = A(t)$), as define by Kenyon in ref. (4), the combination of Eqs. (1) and (2) determine the spatial distribution and temporal evolution of the pressure and deformation fields. The boundary conditions consist of no displacement at the inlet, and no contact stress at the outlet. The pressure is set to $P(0,t) = P_M$ starting from $t = 0$. Given that the fluid and solid are incompressible, the change in outlet pressure (*i.e.,* in the air cavity) is dictated by the sum of the filtration velocity $W$ and the solid deformation temporal gradient $\partial U/\partial t$. Injecting Darcy's law and defining $V$ and $P_A$ as the respective volume and pressure of the cavity, we obtain:

$$\begin{cases} U(0,t) = 0, \\ \dfrac{\partial U}{\partial x}(H,t) = 0, \\ \dfrac{\partial P}{\partial t}(H,t) = \dfrac{\sigma P_A}{V}\left(-\dfrac{\kappa}{\mu}\dfrac{\partial P}{\partial x}(H,t) + \dfrac{\partial U}{\partial t}(H,t)\right), \end{cases} \quad (3)$$

where $\sigma = 15$ mm$^2$ is the material cross section and $H = 0.8$ mm. The analytical solution of Eqs. (1) – (3) for the pressure kinetic response at the outlet position is (see details in Supplementary Material)

$$P(H,t) = P_M\left(1 - 2\sum_{n=1}^{\infty}\frac{\sin(2\lambda_n)\,e^{-\frac{\lambda_n^2 t}{\tau}}}{\sin(2\lambda_n) + 2\lambda_n}\right), \quad (4)$$

where $\tau = \mu H^2/\kappa M$ and $\lambda_n$ are the roots of the constitute relation $\alpha\lambda_n\tan(\lambda_n) = 1$ with $\lambda_n < \lambda_{n+1}$ and $\alpha = VM/\sigma P_A H$. Given that $U(H) = P_M H/2M$ and $W = \kappa P_M/\mu H$ at steady-state, the characteristic time $\tau = 2U(H)/W$ represents the ratio between twice the displacement and the filtration velocity, while $\alpha$ is the ratio of the inlet pressure to the pressure change induced by the solid deformation.

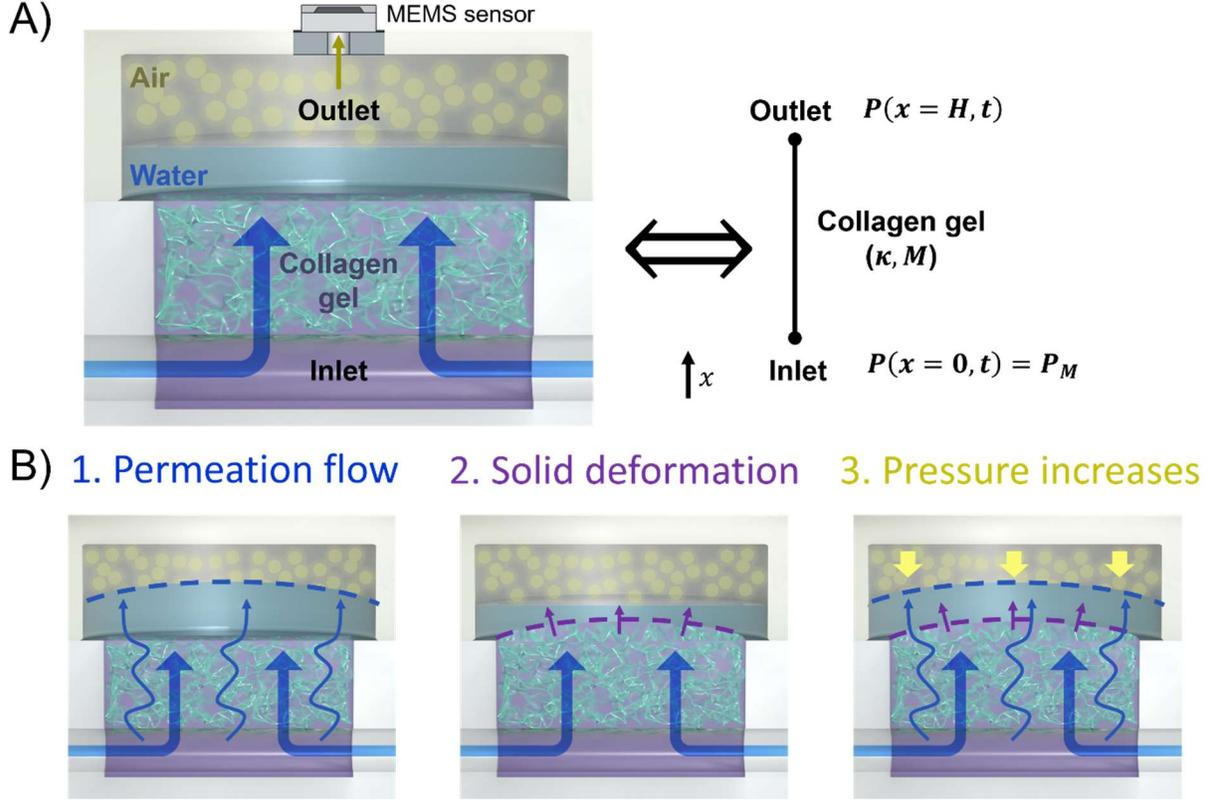

**Fig. 1. Pressure sensing for poroelastic characterization.** A) Principle of the technology: a compartment filled with collagen gel is traversed by an inlet hollow tube of 200 μm in diameter, in which a pressure $P_M$ can be applied. The collagen slab is clamped at the bottom and on the side walls. On top, a MEMS-based pressure sensor records the pressure variation of the intracavity air $P(x = H, t)$. B) A hydrostatic pressure load in the inlet induces a permeation flow and solid deformation that increase the pressure in the cavity.

For a thin and rigid porous material, the contribution of the solid deformation to the cavity pressurization can be neglected, the parameter $\alpha$ becomes much greater than unity ($\alpha \gg 1$) and therefore $\lambda_n \approx 1/\sqrt{\alpha}$ and $\sin(2\lambda_n) \approx 2\lambda_n$. After inserting these results into Eq. (4), the following mono-exponential relaxation for the intracavity pressure is obtained

$$P(H, t) = P_M \left(1 - e^{-\frac{t}{\alpha\tau}}\right). \tag{5}$$

Equation (5) can also be obtained by neglecting the deformation of the solid in Eq. (3) and assuming a linear pressure gradient. This mono-exponential regime can readily be tested using an isoporous polycarbonate membrane (IT4IP) with homogeneous pores of radius $r_p = 0.9$ μm and density $\rho = 3.6 \times 10^5$ cm$^{-2}$. According to these specifications, the permeability of this membrane is given by $\rho \pi r_p^4 / 8$ and is therefore equal to $9.3 \times 10^{-16}$ m$^2$. Considering that the elastic modulus of polycarbonate ~2 GPa (*18*) and the thickness of the membrane ~50 μm, we conclude that $\alpha \sim 10^8$, for a cavity of volume $V = 2$ mL. Hence, the cavity pressurization is well determined by the permeation, and the pressure kinetic data accordingly follow an exponential relaxation (red circles and solid line in Fig. 2A). The fit with Eq. (5) yields a permeability of $(8.7\pm1.1)\times10^{-16}$ m$^2$, which is in excellent agreement with the value inferred from the product datasheet.

Next, we investigate the model predictions for soft poroelastic materials. The evolution of pressure $P(H,t)$ with time is shown in Fig. 2A, for a representative set of permeabilities ($10^{-13} - 10^{-15}$ m$^2$) and elastic moduli (500 – 8000 Pa). In contrast to the exponential relaxation of rigid porous materials (black solid curves), the deformation of the solid speeds up the compression of the cavity in the short-time limit, such that the softer the material, the faster the initial pressure increase. Conversely, the typical relaxation rate in the long-time limit is readily dictated by the permeability, as shown by the clear separation of the relaxation curves for $10^{-13}$, $10^{-14}$, and $10^{-15}$ m$^2$. The highest sensitivity of the pressure model to $M$ and $\kappa$ therefore appears at different time scales, as confirmed by the sensitivity analysis developed in the Supplemental Materials. This fact indicates that both parameters, $M$ and $\kappa$, can accurately be retrieved from relaxation data.

Finally, we test whether the 1D model is able to describe the poroelastic response of a slab crossed by a tubular inlet (see insets of Fig. 2B). We compare the readout of the model to the results of finite element simulations performed with the geometrical settings of our experiments. Sweeping $M$ from 100 to 14000 Pa and $\kappa$ from $0.5\times10^{-14}$ to $16\times10^{-14}$ m$^2$, we extract the deformation of the solid and velocity of the permeation flow at steady-state (red and black datasets in Fig. 2B). These data are proportional to the predictions of the 1D model of $U(H) = P_M H/2M$ and $W = \kappa P_M/\mu H$ with a proportionality factor on the order of unity of 1.31 and 1.77, respectively. The difference of these quantities from unity is mainly because we overlooked the tubular geometry of the inlet in the 1D model. Indeed, the linear gradient in pressure predicted by the 1D model is slower than that in the semi-cylindrical geometry of our system. Qualitatively, we suggest to correct our data by measuring the pressure midway between the inlet and outlet in the simulation, which is equal to $0.69 \times P_M/2$. By multiplying the inlet pressure by this correction factor of 0.69, the 1D model enables us to determine the elastic module and permeability with systematic error factors of 0.90 and 1.22, respectively.

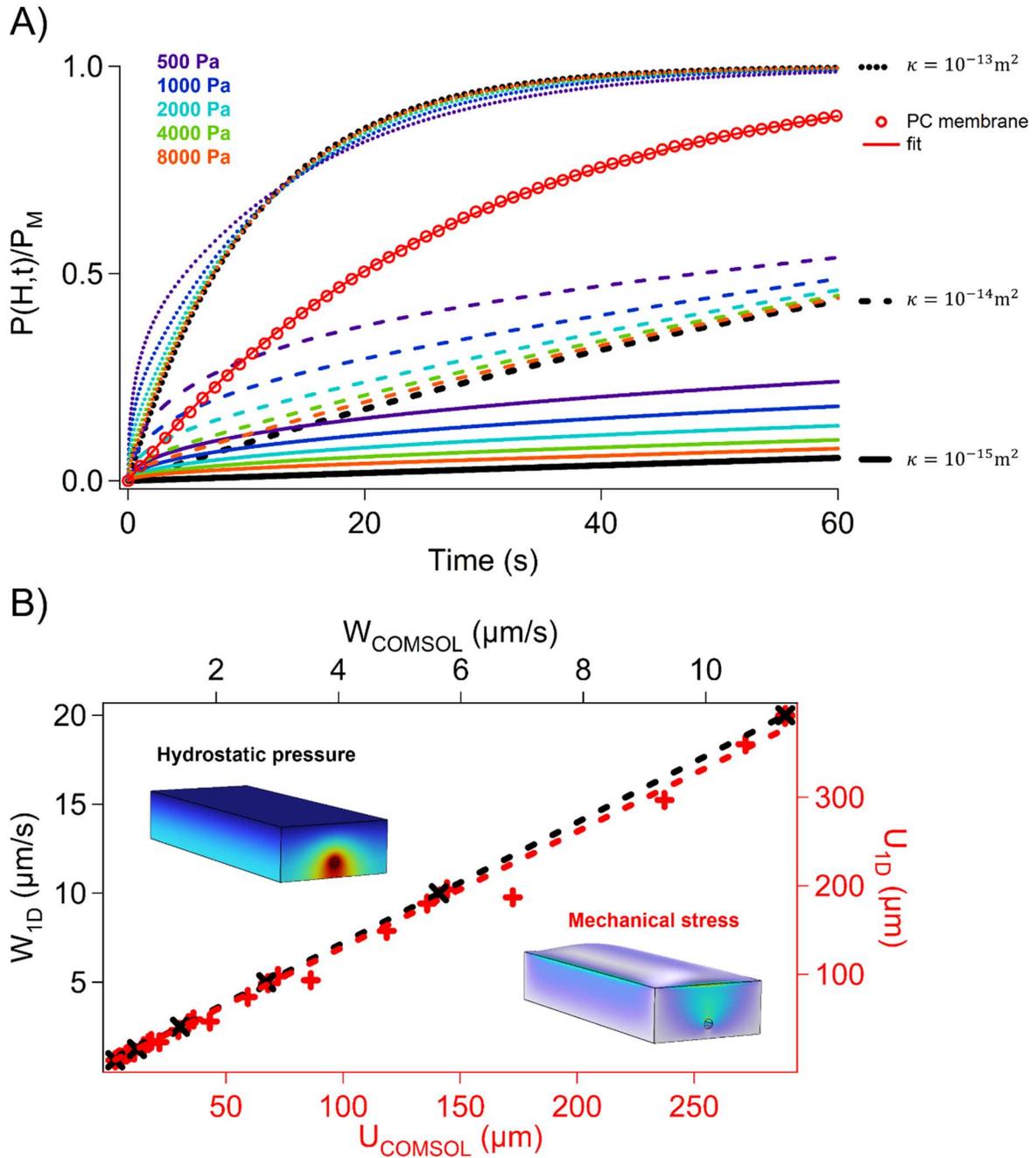

**Fig. 2. Comparison of the predictions of the 1D model and finite element simulations.** A) Intracavity pressure kinetic response predicted the by the 1D model for a P-wave modulus $M \in [500, 8000]$ Pa and three representative values ($10^{-13}$, $10^{-14}$, $10^{-15}$ m$^2$) of the permeability $\kappa$ for a given height of 0.8 mm. The black curves correspond to the limit of a rigid porous material described by Eq. (5) and the red circles to the temporal variation of the intracavity pressure recorded with an isoporous membrane and its fitting (solid red line) via Eq. (5). B) Steady-state fluid velocity and deformation at $x = H$ predicted by the 1D model as functions of the 3D output of finite element simulations performed in COMSOL Multiphysics 6.0. Calculations were done for an air cavity of 2 mL and a porous material of 0.8 mm in height.

*II. Poroelastic characterization of collagen gels*

We prepared collagen gels at a concentration of 2.4 mg/mL using thermal gelation at 37°C during 45 minutes using the protocol described in ref. (*19*). Collagen gels were either used in their native conformation or cross-linked (CL) with para-formaldehyde at a concentration of 4% for 10 hours.(*20*) The pressure response in the cavity was recorded for actuation pressures in the range of 100 to 1000 Pa (red and blue datasets with positive values in Fig. 3A-B). In these settings, collagen properties are probed in tension because the material bulges and expands in the cavity due hydrostatic stress (upper inset in Fig. 3A). After waiting for the system to equilibrate, *i.e.,* reach balanced pressure in the inlet and outlet, we released the pressure at the inlet, and recorded the decrease of intra-cavity pressure. Because the pressure in the cavity is greater than that in the inlet, collagen is in compression (lower inset in Fig. 3A, and curves in the light grey sectors of Fig. 3A-B). These experiments indicate a marked slowdown of the relaxation dynamics in compression for native *vs.* CL collagen gels, whereas the responses in tension are comparable. The fit of these datasets to infer $M$ and $\kappa$ (Fig. 3C-D) explains that these slow kinetics in native collagen networks by the reduced permeability in compression by a factor of four (Fig. 3D). This asymmetric response of native collagen gel in compression *vs.* tension, which is reminiscent of the exponential reduction of cartilage permeability under compression (*21*), became roughly symmetrical after CL. Similarly, the strain-stiffening response of the network, an effect thoroughly described in the literature (*8*), was strongly and mildly asymmetrical in tension *vs.* compression for native and CL collagen, respectively (Fig. 3C). Notably, the tensile elastic modulus increased linearly with the applied stress, in agreement with the data of ref. (*8–10*), and its amplitude was typically two times larger for CL collagen gels. Conversely, the compressive elastic modulus of native collagen became stiffer than that of CL collagen at high load whereas it was softer below 200 Pa. Altogether, native collagen gel is characterized by strain-dependent permeability and asymmetrical strain-stiffening in tension and compression, both effects being reduced or removed by chemical CL.

To investigate the origin of these asymmetric properties, we characterized the microstructure of collagen networks by immunofluorescence confocal microscopy focusing right on top of the pressure inlet. The relaxed structure (central panel, Fig. 3E) showed that the fixation process induced the collapse of the dense native network onto a "primitive" backbone with an increase in pore size. The loose meshwork of CL collagen gels remained stable under tensile and compressive stresses of 100 Pa, as well as under larger loads of 300 Pa (data not shown). This observation was consistent with the steady value of the permeability of this material under tension and compression. Contrariwise, the microstructure of native collagen gels markedly changed in compression, as the fibrous network could no longer be resolved by confocal microscopy (left panel in Fig. 3E). We could not detect structural differences in the meshwork when collagen was in tension (right panel, Fig. 3E). We suggest that the denser meshwork in compression is associated to enhanced hydrodynamic interactions within the pores, and accounts for the strain-dependent permeability. We conclude that microstructural analysis supports the asymmetrical permeability of collagen gels detected by pressure measurements.

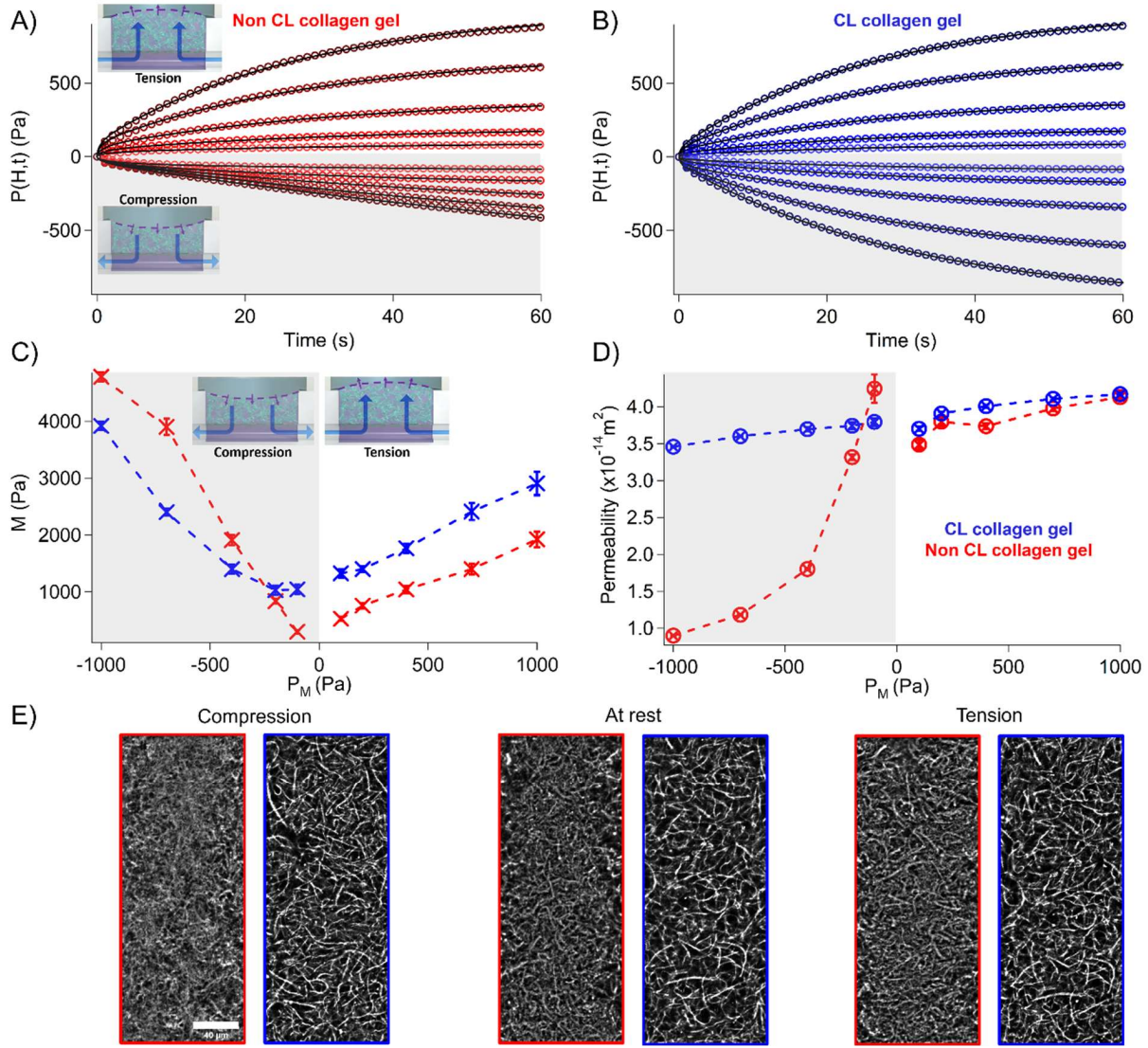

**Fig. 3. Poroelastic analysis of type I collagen gel.** A) before and B) after chemical CL for a pressure $P_M$ ranging from 100 to 1000 Pa. The fitted parameters $M$ and $\kappa$ are plotted as functions of the actuation pressure $P_M$ in panels C) and D), respectively. The dataset for native and CL collagen are reported in red and blue, respectively. E) Representative confocal micrographs of native (in red) and CL (in blue) collagen microstructure under compression (−100 Pa), at rest, and under tension (+100 Pa). The confocal micrograph thickness is 2 μm, as obtained with a 40× water immersion objective. The scale bar corresponds to 40 μm.

*III. Periodic pressure actuation*

Capillary blood pressure consists of a continuous component of ~2000 Pa and periodic pulses, which are characterized by a typical frequency $\omega_0$ of 1 Hz and a magnitude of 100 – 1000 Pa (*22*). We investigated the consequences of this periodic component using a sinusoidal input pressure $P_M$ of amplitude 200, 400, and 800 Pa. After a transient of a few cycles, a permanent regime of oscillations was reached in the cavity, allowing us to compute the pressure averaged over 100 cycles (Fig. 4A-B). The amplitude of the signal in the cavity was ~4% of the input, equivalently

~10 Pa. Using CL collagen gels, the pressure in the cavity had symmetric positive and negative half cycles, was centered at $P_M/2$, and its amplitude proportional to $P_M$ (Fig. 4B). By contrast, native collagen network response markedly changed as the pressure increased. For 200 Pa, the signal was symmetrical and centered at $P_M/2$, but positive and negative half cycles became increasingly asymmetrical as the pressure increased from 400 to 800 Pa (Fig. 4A). This response was associated to an increase of the mean pressure in the cavity to $0.67 \times P_M$ at 800 Pa (inset of Fig. 4A). Hence, periodic pressure was associated to an accumulation of compressive stress in native, but not CL, collagen gels at steady state, and set this material in the regime of sharp variation of its permeability (Fig. 3D).

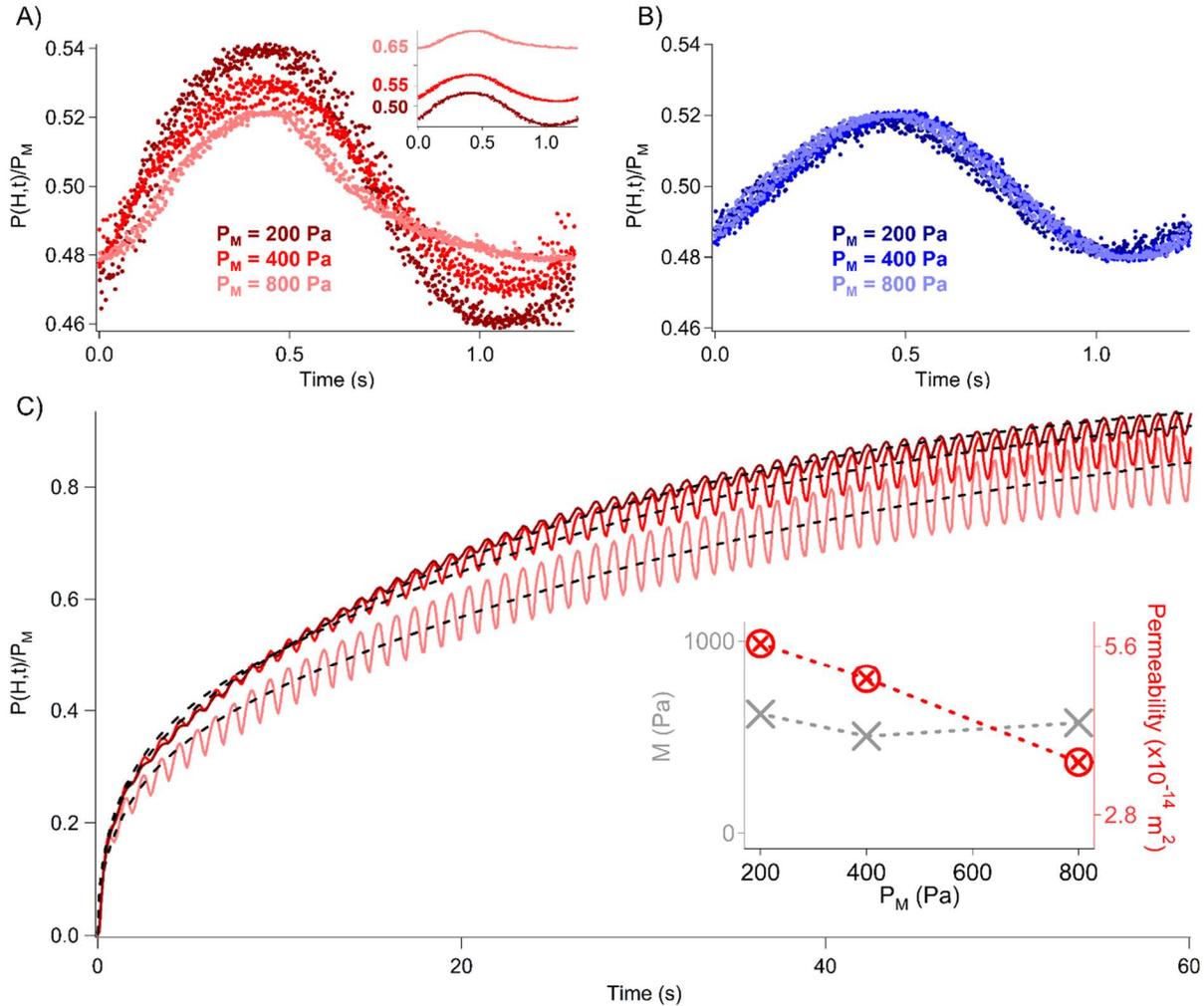

**Fig. 4. Collagen gel response to periodic actuation.** A) The intracavity pressure averaged over 100 cycles for native collagen is normalized and registered at 0.5 for different pressure settings. The inset shows the same raw datasets without registration at 0.5. B) The same experiment is carried out with CL collagen, and the three recordings are only normalized to the input pressure. C) The response of collagen to a step-in pressure is dependent on the magnitude of the amplitude of periodic pressure actuation. The recordings in dark red, red, and light red correspond to an actuation pressure of 200, 400, 800 Pa, respectively. The dashed lines are the fits with the model in Eq. (4) and fitted parameters, $M$ and $\kappa$, are plotted in the inset.

The accumulation of compressive stress can readily be understood from the elegant analysis of pressure consolidation in poroelastic materials proposed by Kenyon (23). Fluid injection with a step in pressure is associated to a diffusion-like propagation of the pressure field in the material (17). The effective diffusion coefficient $D$ scales as $\kappa M/\mu$, resulting in a consolidation layer of comparable thickness in tension and in compression of $\sqrt{D \times \pi/\omega_0}$ ~140 µm. Liquid is hence soaked into and expelled from the collagen gel during the cycles of tension and compression, respectively. The flow velocity during each cycle is typically determined by the pressure gradient in the consolidation layer and the permeability:

$$W \sim \frac{\kappa}{\mu} \frac{P_M}{\sqrt{D \times \pi/\omega_0}} = \sqrt{\frac{\kappa}{\mu M}} \frac{P_M}{\sqrt{\pi/\omega_0}} \qquad (6)$$

The asymmetric response of collagen is characterized by a reduced permeability and an increased elasticity in compression by the same typical factor of four. Hence, the velocity $W$ is reduced by a factor of four in compression vs. tension. Mass conservation implies that the same amount of fluid is expelled and soaked during compressive and tensile cycles. This condition is fulfilled if the soaking phase is shorter than that of expulsion, as insured by the offset in bulk compressive stress. We note that this offset in compressive stress enables the dynamic modulation of collagen gel permeation properties. This is shown by monitoring the kinetic response to a step of pressure of 500 Pa with a permanent periodic pulsatile flow associated to an amplitude 100, 200, and 400 Pa (dark, plain, and light red curves in Fig. 4C). The relaxation became slower as the amplitude of the periodic stimulation increased, in agreement with the fact that periodic forcing set collagen in conditions of low permeability. The fit of the curves with the model of Eq. (4) (dashed black curves in Fig. 4C) confirmed our hypothesis because the permeability was 35% lower for high amplitude pulsations and the elastic modulus was unchanged (inset of Fig. 4C).

**Discussion**

We have developed a methodology and its analytical model for the characterization of virtually any poroelastic material. Applied to type I collagen gels, the method allows us to show that this material is characterized by enhanced strain-stiffening and reduced permeability in compression. This asymmetrical property can have profound consequences on the transport in biological tissues in the context of the extended Starling principle (24). Considering the "healthy" situation of a steady interstitial flow between blood and lymphatic microvessels, pulsatile intraluminal pressure in the blood circulation network creates a compressive stress that tends to reduce permeation flux. Lymphatic drainage vanishes this stress, likely facilitating clearance in the tissue by enhancing the permeability (25). Upon physical efforts, the increase of the blood pressure is associated to high amplitude pulsatile flows that are likely to reduce the permeability of the tissue, hence hinting to a passive regulatory role of collagen gels to control homeostasis. If we now consider the pathological environment of tumors, we suggest that the mechanical compression induced by the tumor mass cannot be reset by the lymphatic system, reducing drainage and enhancing hypoxia (26). This effect may be further enhanced by tissue remodeling in the tumor environment (27), because the enriched concentration of collagen is expected to lower the tissue permeability. Consequently, poroelastic properties of tissues and their alteration during tissue remodelling by *e.g.*, tumor growth appears as an underestimated player of homeostatic regulation, which could be studied and quantified with microphysiological systems.

**Materials and methods**
Collagen gel chips were fabricated using the microvessel chips described in ref. (*19*). The silicone chips and 200 μm (No. 08, J type; Seirin, Shizuoka, Japan) were first placed together in a vacuum reactor with 100 μL of aminopropyl-triethoxysilane, and left at 0.1 mbar and room temperature for 30 minutes. Needles were then soaked in 1% (m:v) bovine serum albumin, dried, and sterilized by UV-light exposure. The chips were treated with 50 μL of 2.5% glutaraldehyde for 1 minute, then thoroughly rinsed with water, and dried. The collagen solution was subsequently prepared on ice by mixing Cellmatrix Type I-P collagen solution (Nitta Gelatin, Japan), 10× Hanks' buffer, and 10× collagen buffer (volume ratio 8:1:1) following manufacturer's protocol. We poured 30 μL of this ice-cold collagen solution into the chip, inserted the coated needle, and incubated the system at 37°C for 40 min to allow collagen reticulation. The needle was withdrawn to form a hollow channel, and the chips were left in phosphate bufffer.

The collagen gel embedded within a silicone chip was connected to a pressure controller (Fluigent MFCS) and a pressure sensor (Merit® Sensor LP series) by mean of 3D printed mechanical pieces (Elegoo Mars 3, water washable resin). Acquisition was done through a dedicated LabVIEW interface allowing the control of the inlet pressure while recording the outlet signal with a digital multimeter (Agilent, 34401A).

Simulations were run with COMSOL Multiphysics 6.0 using the linear poroelastic module, which allowed coupling of the Darcy Law and the Solid Mechanics modules. We set the tube radius $r_0$ to 100 μm and placed it 300 μm above the glass coverslip. The collagen slab was 2.5 mm in width and 0.8 mm in height. An inlet pressure $P_M$ was imposed onto the collagen gel channel border, creating a double boundary condition defined as a pressure in the Darcy Law module and a Boundary load in the Solid mechanics module. The outlet pressure was imposed to 0 Pa onto the top collagen gel border.

Immunofluorescence confocal microscopy was performed with the LSM 700 confocal microscope (Carl Zeiss) equipped with a 40× water immersion objective (numerical aperture of 1.2), and using a pinhole of 1 Airy unit. After chemical cross-linking of collagen gels and blocking with bovine serum albumin overnight, we incubated a collagen mouse primary antibody (Cell Signaling Technology) complexed with a secondary antibody (1:400, 488 nm) for 4 hours, and then rinsed the chip with phosphate buffer. The chip was mounted on the pressure sensing system, opening the pressure sensor outlet to apply a steady pressure gradient over time.

**Acknowledgment**
The authors thank Takashi Ando for his artwork and Rachel Tan for English proofreading.

**Funding:** JC acknowledges the JSPS for a postdoctoral fellowship. The authors thank the LIMMS (CNRS-Institute of Industrial Science, University of Tokyo) for financial support. This research was partly supported by the Grant-in-Aid for JSPS Fellows (20F20806), AMED P-CREATE (JP18cm0106239h0001), and JSPS Core-to-Core Program (JPJSCCA-20190006).

**Author contribution:**
Conceptualization: JC, AB, JOM
Methodology: JC, AB, JOM
Investigation: JC
Software: LJ
Visualization: JC
Supervision: JC, AB, YTM
Writing—original draft: JC, AB
Writing—review & editing: JOM, YTM, MN, LJ

**Competing interests:** Authors declare that they have no competing interests.

**Data and materials availability:** All data are available in the main text or the supplementary materials.

**Asymmetry of tensile *vs*. compressive elasticity and permeability contributes to the regulation of exchanges in collagen gels**

Jean Cacheux,[1,2] Jose Ordonez-Miranda,[1,2] Aurélien Bancaud,[1,2,3]* Laurent Jalabert,[1,2] Masahiro Nomura,[1,2] Yukiko T. Matsunaga[1,2]†

[1] *Institute of Industrial Science, The University of Tokyo, Tokyo 153-8505, Japan.*
[2] *LIMMS, CNRS-IIS IRL 2820, The University of Tokyo, Tokyo 153-8505, Japan.*
[3] *LAAS-CNRS, Université de Toulouse, CNRS, Toulouse, France.*
*abancaud@laas.fr; †mat@iis.u-tokyo.ac.jp

**Supplementary Materials**

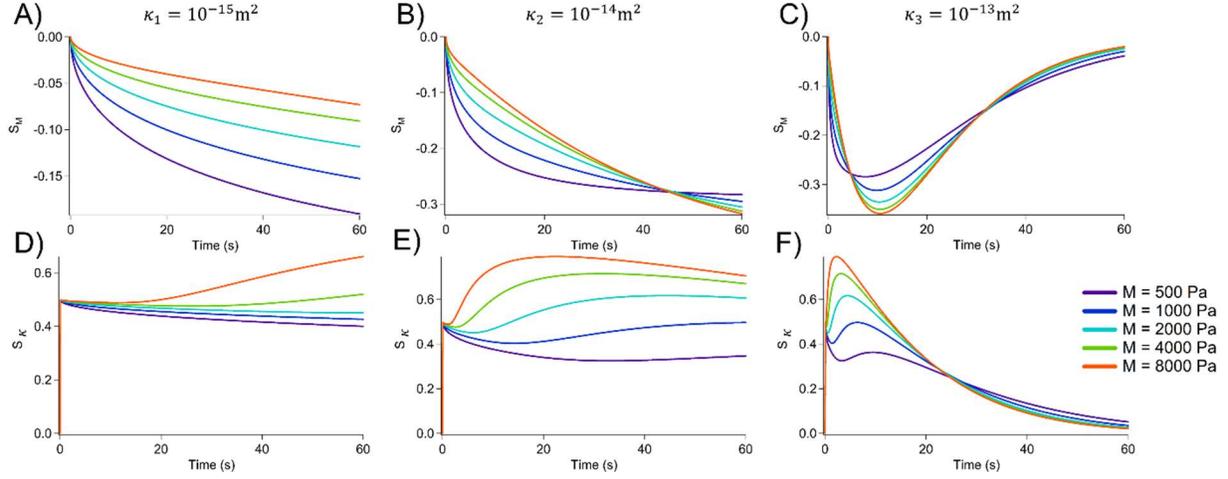

**Fig. S1.** We plot A-C) $S_M = \frac{\partial P(H,t)}{\partial M}\frac{M}{P(H,t)}$ and D-F) $S_\kappa = \frac{\partial P(H,t)}{\partial \kappa}\frac{\kappa}{P(H,t)}$ for $\kappa$ ranging from $10^{-15}$ to $10^{-13}$ m$^2$ and $M$ from 500 Pa to 8000 Pa.

## 1D model: Laplace transform

The flux and equilibrium equations, Eq. (1), Eq. (2) and Eq. (3), can be combined to give:

$$\frac{\kappa}{\mu}M\frac{\partial^2 U}{\partial x^2} = \frac{\partial U}{\partial t} - A(t) ;$$

$U(x) = \frac{1}{2\pi}\int_{-\infty}^{+\infty}\bar{U}(x,t)e^{-st}dt$ with $\bar{U}$ the Laplace transform of the deformation.

$$\Rightarrow \frac{\kappa}{\mu}M\frac{\partial^2 \bar{U}}{\partial x^2} = s\bar{U} - \bar{A} ; \delta^2 = \frac{\kappa}{\mu}\frac{M}{s}$$

$$\Rightarrow \delta^2 \frac{\partial^2 \bar{U}}{\partial x^2} - \bar{U} = -\frac{\bar{A}}{s} ; n = \frac{x}{\delta}$$

$$\Rightarrow \frac{\partial^2 \bar{U}}{\partial n^2} - \bar{U} = -\frac{\bar{A}}{s} \Rightarrow \bar{U}(n) = \frac{\bar{A}}{s} + Be^n + Ce^{-n}$$

$$\Rightarrow \frac{d\bar{P}}{dx} = \frac{\mu}{\kappa}(s\bar{U} - \bar{A}) = \frac{s\mu}{\kappa}\left(\bar{U} - \frac{\bar{A}}{s}\right) = \frac{s\mu}{\kappa}(Be^n + Ce^{-n})$$

$$\frac{d\bar{P}}{dn} = \frac{s\mu\delta}{\kappa}(Be^n + Ce^{-n}) \Rightarrow \bar{P}(n) = D + \frac{s\mu\delta}{\kappa}(Be^n - Ce^{-n})$$

<u>Boundary conditions:</u>

① $U(0,t) = 0 \Rightarrow \bar{U}(0) = 0 = \frac{\bar{A}}{s} + \bar{B} + \bar{C}$

② $\frac{\partial U}{\partial x}(H,t) = 0 \Rightarrow \bar{U}'(n_H) = 0 = Be^{n_H} - Ce^{-n_H} ; n_H = \frac{H}{\delta}$

③ $\frac{\partial P}{\partial t}(H,t) = \frac{\sigma P_A}{V}\left(-\frac{\kappa}{\mu}\frac{\partial P}{\partial x}(H,t) + \frac{\partial U}{\partial t}(H,t)\right) \Rightarrow \bar{P}(n_H) = \frac{\sigma P_A}{V}\left(\bar{U}(n_H) - \frac{\kappa}{s\mu\delta}\bar{P}'(n_H)\right)$

$$\bar{P}(n_H) = \frac{\sigma P_A}{V}\left(\frac{\bar{A}}{s} + Be^n + Ce^{-n} - (Be^n + Ce^{-n})\right) = \frac{\sigma P_A}{V}\frac{\bar{A}}{s}$$

$$D + \frac{s\mu\delta}{\kappa}(Be^{n_H} - Ce^{-n_H}) = \frac{\sigma P_A}{sV}\bar{A} \Rightarrow D = \frac{\sigma P_A}{sV}\bar{A}$$

From boundary condition ②: $B = Ee^{-n_H}$ and $C = Ee^{-n_H}$

From boundary condition ①: $\frac{\bar{A}}{s} + 2E\cosh(n_H) = 0 \Rightarrow \frac{\bar{A}}{s} = -2E\cosh(n_H)$

$$\Rightarrow \bar{P}(n) = D + \frac{s\mu\delta}{\kappa}E(e^{n-n_H} - e^{n_H-n}) = D - \frac{s\mu\delta}{\kappa}2E\sinh(n_H - n)$$

$$\Rightarrow \bar{P}(n) = D + \bar{A}\frac{\mu\delta}{\kappa}\frac{\sinh(n_H-n)}{\cosh(n_H)} = D\left(1 + \frac{s\mu\delta}{\kappa}\frac{V}{\sigma P_A}\frac{\sinh(n_H-n)}{\cosh(n_H)}\right)$$

④ $P(0,t) = P_M \Rightarrow \bar{P}(0) = \frac{P_M}{s} = D\left(1 + \frac{s\mu\delta}{\kappa}\frac{V}{\sigma P_A}\tanh(n_H)\right)$

$$\Rightarrow D = \frac{P_M}{sQ(s)}\ ;\ Q(s) = 1 + \frac{s\mu\delta}{\kappa}\frac{V}{\sigma P_A}\tanh(n_H)\ ;\ \delta = \sqrt{\frac{\kappa M}{\mu s}}\ ;\ n_H = \frac{H}{\delta}$$

$$\Rightarrow \bar{P}(n) = \frac{P_M}{sQ(s)}\left(1 + \frac{s\mu\delta}{\kappa}\frac{V}{\sigma P_A}\frac{\sinh(n_H-n)}{\cosh(n_H)}\right)$$

$$\Rightarrow \bar{P}(n_H) = \frac{P_M}{sQ(s)} = \frac{P_M}{s\left(1+\frac{V}{\sigma P_A}\sqrt{\frac{s\mu M}{\kappa}}\tanh\left(H\sqrt{\frac{s\mu}{\kappa M}}\right)\right)}$$

$$\Rightarrow \bar{P}(n_H) = \frac{P_M}{s(1+\alpha\sqrt{s\tau}\tanh(\sqrt{s\tau}))}\ ;\ \tau = \frac{\mu}{\kappa}\frac{H^2}{M}\ \text{and}\ \alpha = \frac{V}{\sigma P_A H}M$$

$$\Rightarrow P(H,t) = \frac{P_M}{2\pi i}\int_{\gamma-i\infty}^{\gamma+i\infty}\frac{e^{st}ds}{s(1+\alpha\sqrt{s\tau}\tanh(\sqrt{s\tau}))}$$

Poles: $s = 0$ and $1 + \alpha\sqrt{s\tau}\tanh(\sqrt{s\tau}) = 0 \Rightarrow \sqrt{s\tau} = i\lambda_n$ or $s = -\frac{\lambda_n^2}{\tau}\ ;\ n = 1, 2, \ldots$

$$\Rightarrow 1 + \alpha i\lambda_n\tanh(i\lambda_n) = 0 = 1 + \alpha i\lambda_n i\tan(\lambda_n)$$

$$\Rightarrow \alpha\lambda_n\tan(\lambda_n) = 1\ ;\ n = 1, 2, \ldots$$

$$\frac{P(H,t)}{P_M} = \text{Res}(s = 0) + \sum_{n=1}^{\infty}\text{Res}\left(s = -\frac{\lambda_n^2}{\tau}\right)$$

Residues:

- Res$(s = 0) = \lim_{s \to 0}\frac{se^{st}}{s(1+\alpha\sqrt{s\tau}\tanh(\sqrt{s\tau}))} = 1$

- $\text{Res}\left(s = -\frac{\lambda_n^2}{\tau}\right) = \lim_{s \to -\frac{\lambda_n^2}{\tau}} \frac{\left(s+\frac{\lambda_n^2}{\tau}\right)e^{st}}{s(1+\alpha\sqrt{s\tau}\tanh(\sqrt{s\tau}))} = \frac{e^{-\frac{\lambda_n^2 t}{\tau}}}{-\frac{\lambda_n^2}{\tau}} \lim_{s \to -\frac{\lambda_n^2}{\tau}} \frac{\left(s+\frac{\lambda_n^2}{\tau}\right)}{1+\alpha\sqrt{s\tau}\tanh(\sqrt{s\tau})} =$

$\frac{-\tau e^{-\frac{\lambda_n^2 t}{\tau}}}{\lambda_n^2} \lim_{s \to -\frac{\lambda_n^2}{\tau}} \frac{1}{\alpha\frac{d}{ds}(\sqrt{s\tau}\tanh(\sqrt{s\tau}))} = \frac{-\tau e^{-\frac{\lambda_n^2 t}{\tau}}}{\alpha\lambda_n^2} \lim_{s \to -\frac{\lambda_n^2}{\tau}} \frac{1}{(\sqrt{s\tau}\,\text{sech}^2(\sqrt{s\tau})+\tan\,(\sqrt{s\tau}))} 2\sqrt{\frac{s}{\tau}} =$

$\frac{-2}{\alpha\lambda_n} \frac{e^{-\frac{\lambda_n^2 t}{\tau}}}{\lambda_n \sec^2(\lambda_n)+\tan(\lambda_n)} = \frac{-2e^{-\frac{\lambda_n^2 t}{\tau}}}{1+\alpha\lambda_n^2 \sec^2(\lambda_n)}$ because $\tan(\lambda_n) = \frac{1}{\alpha\lambda_n}$

$$\text{Res}\left(s = -\frac{\lambda_n^2}{\tau}\right) = \frac{-2\cos^2(\lambda_n)e^{-\frac{\lambda_n^2 t}{\tau}}}{\alpha\lambda_n^2+\cos^2(\lambda_n)}$$

$\Rightarrow P(H,t) = P_M\left(1 - 2\sum_{n=1}^{\infty} \frac{\cos^2(\lambda_n)e^{-\frac{\lambda_n^2 t}{\tau}}}{\alpha\lambda_n^2+\cos^2(\lambda_n)}\right)$

Where $\lambda_n$ are the roots of $\alpha\lambda_n \tan(\lambda_n) = 1$ with $\lambda_n < \lambda_{n+1}$

Remark:

$\frac{\cos^2(\lambda_n)}{\alpha\lambda_n^2+\cos^2(\lambda_n)} = \frac{\cos^2(\lambda_n)}{\cos^2(\lambda_n)+\lambda_n\frac{\cos(\lambda_n)}{\sin(\lambda_n)}} = \frac{\sin(\lambda_n)\cos(\lambda_n)}{\sin(\lambda_n)\cos(\lambda_n)+2\lambda_n} = \frac{\sin(2\lambda_n)}{\sin(2\lambda_n)+2\lambda_n}$

$P(H,t) = P_M\left(1 - 2\sum_{n=1}^{\infty} \frac{\sin(2\lambda_n)e^{-\frac{\lambda_n^2 t}{\tau}}}{\sin(2\lambda_n)+2\lambda_n}\right)$

### 1D model: temporal sensitivity

#### Partial derivative with respect to $\alpha$

- $\frac{\partial P}{\partial \alpha} = \frac{\partial P}{\partial \lambda_n}\frac{\partial \lambda_n}{\partial \alpha}$ ; $P(H,t) = P_M\left(1 - 2\sum_{n=1}^{\infty} F(\lambda_n)e^{-\frac{\lambda_n^2 t}{\tau}}\right)$

- $\frac{\partial P}{\partial \lambda_n} = -2P_M \sum_{n=1}^{\infty}\left(F' - F\frac{2\lambda_n t}{\tau}\right)e^{-\frac{\lambda_n^2 t}{\tau}}$

  With $F' = \frac{\partial F}{\partial \lambda_n} = \frac{2\cos(2\lambda_n)(\sin(2\lambda_n)+2\lambda_n)-\sin(2\lambda_n)(2+2\cos(2\lambda_n))}{(\sin(2\lambda_n)+2\lambda_n)^2} = \frac{2(2\lambda_n \cos(2\lambda_n)-\sin(2\lambda_n))}{(\sin(2\lambda_n)+2\lambda_n)^2}$

  $\frac{\partial P}{\partial \lambda_n} = 4P_M \sum_{n=1}^{\infty}\left(\frac{(\sin(2\lambda_n)-2\lambda_n\cos(2\lambda_n))}{(\sin(2\lambda_n)+2\lambda_n)^2} + \frac{\sin(2\lambda_n)}{\sin(2\lambda_n)+2\lambda_n}\frac{\lambda_n t}{\tau}\right)e^{-\frac{\lambda_n^2 t}{\tau}}$

- $\frac{\partial \lambda_n}{\partial \alpha}$

  $\alpha\lambda_n \tan(\lambda_n) = 1 \Rightarrow \frac{\partial}{\partial \alpha}(\alpha\lambda_n \tan(\lambda_n)) = 0$

  $\Rightarrow \lambda_n \tan(\lambda_n) + \alpha\frac{\partial}{\partial \lambda_n}(\lambda_n \tan(\lambda_n))\frac{\partial \lambda_n}{\partial \alpha} = 0$

$$\Rightarrow \frac{1}{\alpha} + \alpha(\tan(\lambda_n) + \lambda_n \sec^2(\lambda_n))\frac{\partial \lambda_n}{\partial \alpha} = 0$$

$$\Rightarrow \frac{1}{\alpha} + \alpha\left(\frac{\sin(\lambda_n)}{\cos(\lambda_n)} + \frac{\lambda_n}{\cos^2(\lambda_n)}\right)\frac{\partial \lambda_n}{\partial \alpha} = 0$$

$$\Rightarrow \frac{\cos^2(\lambda_n)}{\alpha} + \alpha(\cos(\lambda_n)\sin(\lambda_n) + \lambda_n)\frac{\partial \lambda_n}{\partial \alpha} = 0$$

$$\Rightarrow \frac{\partial \lambda_n}{\partial \alpha} = \frac{-\alpha^{-1}\cos^2(\lambda_n)}{\alpha(\cos(\lambda_n)\sin(\lambda_n) + \lambda_n)} = \frac{-\cos^2(\lambda_n)}{\alpha^2(\cos(\lambda_n)\sin(\lambda_n) + \lambda_n)} = \frac{-(\lambda_n \sin(\lambda_n))^2}{\cos(\lambda_n)\sin(\lambda_n) + \lambda_n} = \frac{-2\lambda_n^2 \sin^2(\lambda_n)}{\sin(2\lambda_n) + 2\lambda_n}$$

Because $\cos(\lambda_n)\sin(\lambda_n) = \frac{\sin(2\lambda_n)}{2}$

$$\frac{\partial P}{\partial \alpha} = -8P_M \sum_{n=1}^{\infty}\left(\sin(2\lambda_n) - 2\lambda_n \cos(2\lambda_n) + (\sin(2\lambda_n) + 2\lambda_n)\sin(2\lambda_n)\frac{\lambda_n t}{\tau}\right)\frac{\lambda_n^2 \sin^2(\lambda_n)}{(\sin(2\lambda_n) + 2\lambda_n)^3}e^{-\frac{\lambda_n^2 t}{\tau}}$$

**Partial derivative with respect to $\tau$**

$$\frac{\partial P}{\partial \tau} = -2P_M \sum_{n=1}^{\infty}\frac{\lambda_n^2 t}{\tau^2}\frac{\sin(2\lambda_n)e^{-\frac{\lambda_n^2 t}{\tau}}}{\sin(2\lambda_n) + 2\lambda_n}$$

**Partial derivative with respect to $M$**

$$\frac{\partial P}{\partial M} = \frac{\partial P}{\partial \alpha}\frac{\partial \alpha}{\partial M} + \frac{\partial P}{\partial \tau}\frac{\partial \tau}{\partial M} \Rightarrow \frac{\partial P}{\partial M} = \frac{V}{\sigma P_A H}\frac{\partial P}{\partial \alpha} - \frac{\mu H^2}{\kappa M^2}\frac{\partial P}{\partial \tau}$$

**Partial derivative with respect to $\kappa$**

$$\frac{\partial P}{\partial \kappa} = \frac{\partial P}{\partial \alpha}\frac{\partial \alpha}{\partial \kappa} + \frac{\partial P}{\partial \tau}\frac{\partial \tau}{\partial \kappa} \Rightarrow \frac{\partial P}{\partial \kappa} = \frac{\partial P}{\partial \tau}\frac{\partial \tau}{\partial \kappa} = -\frac{\mu H^2}{\kappa^2 M}\frac{\partial P}{\partial \tau}$$